\documentclass{caosp307}

\usepackage{graphicx}

\usepackage{csquotes}

\usepackage{natbib}
\articleNo{295}
\pubyear{2020}
\volume{50}
\volnumber{4}
\firstpage{717}
\received{April 13, 2020}
\accepted{June 9, 2020}

\begin{document}

\hauthor{M.Yu.\,Skulskyy}

\title{Formation of magnetized spatial structures in the Beta Lyrae system }
\subtitle{ II. Reflection of magnetically controlled structures in the visible spectrum}

\author{
M.Yu.\,Skulskyy
       }

\institute{
Lviv Polytechnic National University, Department of Physics, 79013, Lviv, Ukraine, \email{mysky@polynet.lviv.ua}
          }

\date{April 13, 2020}

\maketitle

\begin{abstract}
This article proposes a picture of magnetized accretion structures formed during the mass transfer in the Beta Lyrae system. It is shown that the structure of the gaseous flows between the donor and the gainer is due to the spatial configuration of the donor magnetic field. Its dipole axis is deviated substantially from the line joining the centers of the components and is inclined to the orbital plane of the binary system; the center of the magnetic dipole is displaced from the donor center toward the gainer. The surface around the donor magnetic pole, which is close to the gainer, is a region of an additional matter loss from the donor surface. The effective collision of the magnetized plasma with the accretion disk is enhanced by the fast counter-rotation of this disk, especially in the secondary quadrature phases, in which the high-temperature medium and the system of formed accretion flows are observed.

This concept is demonstrated, primarily, in the obvious correlations between the phase variability of the donor magnetic field and the corresponding variability of the dynamic and energy characteristics of the various complex lines. This refers to the behavior of the radial velocity curves of the emission-absorption lines formed in the gaseous structures of type H$_{\alpha}$, He\,I $\lambda$ 7065, or the variability of their equivalent width and intensity, and the variability of conventional absorption lines of the donor atmosphere. This is true for the phase variability of the absolute flux in the H$_{\alpha}$ emission line and the fast varying of the continuum in the H$_{\alpha}$ region as certain parameters, which reflect the phase variability of the donor magnetic field. This approach made it possible to determine the phase boundaries of the location of the magnetic polar region on the donor surface above which the matter outflows are formed.
\end{abstract}

\section{Introduction}
\label{intr}
The previous article \citep{Skulskyy2020} was intended to provide an overview and analysis of long-term observations needed to further highlight the questions aimed primarily on the study of magnetized gaseous structures. This thematics was decisive and intensively developed on the basis of spectral observations on large telescopes with the latest equipment in 1980-1995. Discoveries made in original investigations, such as \cite{Skulskij1985,Skulskij1992,Alekseev1989,Burnashev1991,Skulskii1992-Malkov,Skulsky1993}, and others, have allowed us to create a certain picture of magnetized gaseous structures until the mid-1990s. Along with a number of such studies over the next decade, this picture has changed little. The following points should be briefly noted. The analysis of all observations and studies of the donor magnetic field showed that the systematic observations on the 6-m SAO telescope could be considered decisive for the further study of the mass transfer and the formation of accretion structures in the Beta Lyrae system. Based on these observations, in \cite{Skulskij1985} the first simulation of the donor magnetic field configuration was conducted. The axis of the magnetic field is directed by the orbital phases of (0.355-0.855)\,P. It is also important that the magnetic dipole axis is inclined to the orbit plane of the binary system by an angle of 28$\degr$, and the center of the magnetic dipole is displaced by 0.08 of the distance between centers of gravity of both components from the donor center toward the gainer. It could be assumed that the mass loss and its transfer from the donor to the gainer can occur not only in the direction of the (0.5-1.0)\,P phases of the star-components gravity axis, but also in the direction of the magnetic field axis. That is, in addition to the gas flow that is directed from the deformed donor through a Lagrange point to the gainer's Roche cavity in a classical hydrodynamic picture \cite[see][]{Bisikalo2000}, there is the matter outflow channeled by the donor magnetic field in the direction of its dipole axis from the donor surface and deflects along the magnetic field lines toward the accretion disk. Moreover, the donor magnetic pole in the 0.855\,P phase is located on the donor surface slightly above the orbital plane and approximated to the gainer. This presumes also the presence of more effective shock collisions of the magnetized plasma in the phases of the second quadrature (0.60-0.85)\,P at all heights of the accretion disk. The energy effect of shock collisions of magnetized gas with the accretion disk is amplified in these phases due to the rapid counter-rotation of this disk forming a hot arc on the outer rim of the accretion disk facing the donor, \cite[see][]{Skulsky2015, Skulsky2018}.

These inferences, based on the spatial configuration of the donor magnetic field, required a number of diverse observations and developed gradually. At the same time, other scientists have also pointed out the need for such research. \cite{Bahyl1986}, after having carefully studied the spectral absorption lines of the donor atmosphere, noted: \textquote{the curve of the system’s magnetic field variations is similar in shape to the curves of the phase variations of the equivalent width of the corresponding lines}. \cite{Aydin1988} also indicated that the presence in the Beta Lyrae spectrum of the high-temperature resonance lines of N\,V, Si\,IV, and C\,IV implies the existence of non-thermal sources in this system, which cannot be matched by the radiation from stellar components; however, this can be considered in the context of the discovery of a variable magnetic field by \cite{Skulskij1985}. Recently, \cite{Ignace2018}, while studying the phase variability of complex emission-absorption profiles of the H$_\alpha$ line, concluded that in the explanation of the observation data \textquote{the magnetic field...could prove relevant, e.g., \cite{Skulsky2018}}. 

Our analysis of known published observations has shown that the donor magnetic field is in some way reflected in infrared, optical and UV spectral regions. This has stimulated the detailed investigation of the Beta Lyrae spectrum to present it in the summarized work. This article focuses mainly on the extended study of the continuum and complex lines in the visual spectrum with a view of detecting, understanding, and explanation of the relationships between the phase variability of the donor magnetic field and the characteristics of the physical processes occurring in moving gaseous structures between the donor and the gainer.

It should be recalled that the longitudinal component of the donor's magnetic field changes significantly during the orbital period (which is close to 12.94\,d); phase changes in the range from zero to one, being tied to the main minimum of this binary system in the visible spectral region when the more massive accretor obscures of the bright donor; the schematic model of the Beta Lyrae system and the picture of mass transfer are shown in \cite{Skulskyy2020} in Figure 1, which is needed to understand the physical processes in further analysis. 

\section{Magnetic field and physical processes in gaseous structures between donor and gainer}
\subsection{Magnetic field and the absolute flux in H$_\alpha$ emission line}
\label{sec:2_1}
\begin{figure}[!t]
	\centerline{\includegraphics[width=0.75\textwidth,clip=]{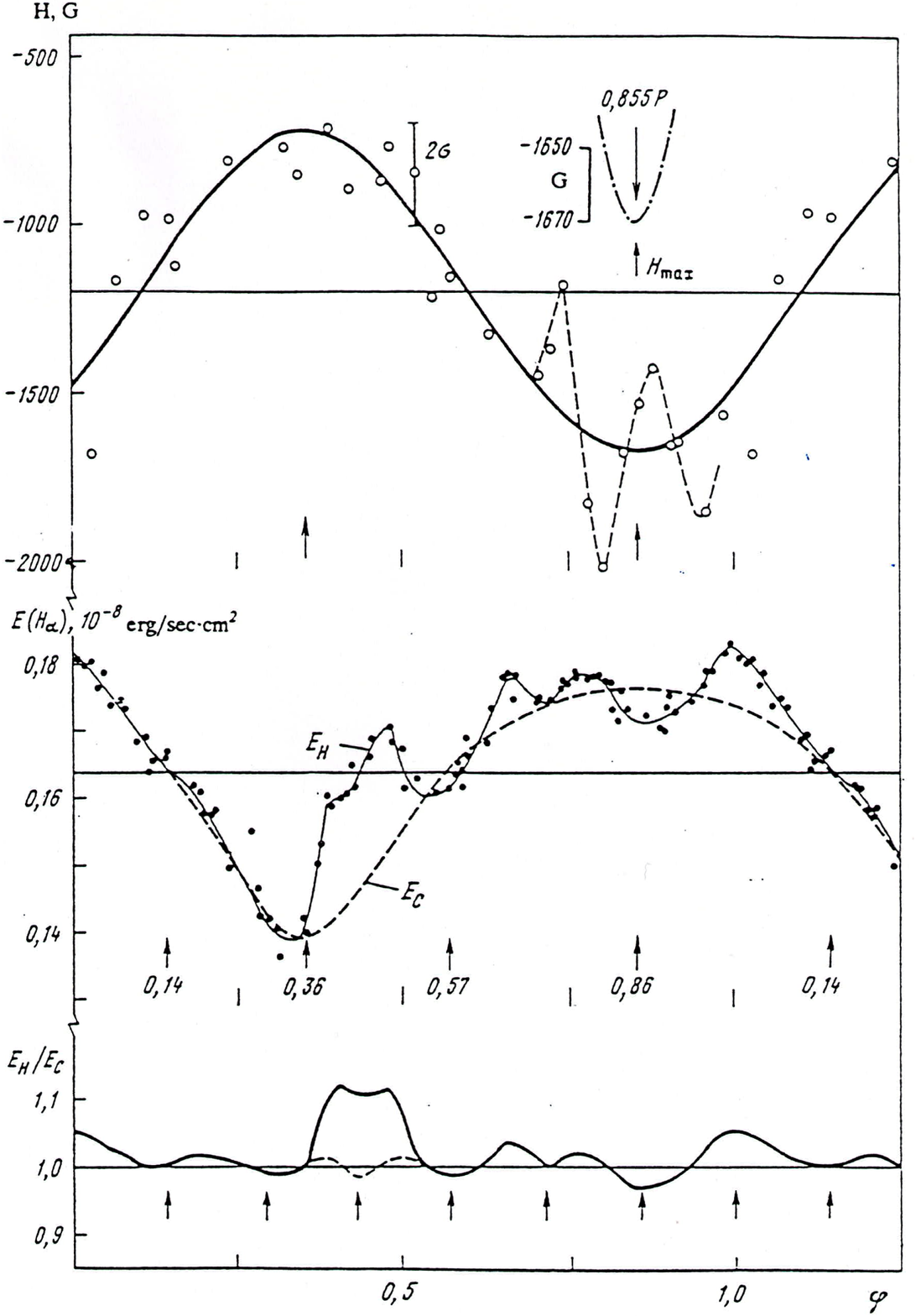}}
	\caption{At the top - the variability over the orbital phase of the effective magnetic field strength of the donor; in the middle - the variability over the orbital phase of the absolute radiation flux in the H$_\alpha$ emission line; bottom - the 12\% increase in the radiation flux in the H$_\alpha$  emission line in the phases round 0.43\,P, which is interpreted as a result shock collisions of gaseous flows with the accretion disk. \citep{Burnashev1991}.}
	\label{fig:1}
\end{figure}

The first among the most important results on the radiation of accretion flows in the Beta Lyrae system was obtained by \cite{Burnashev1991} on the basis of their spectrophotometry that was carried out in 1974-1985. The narrowband photometry was given in values of the monochromatic illuminance at the top of the Earth's atmosphere $\log E(\lambda)$ in $erg \cdot sec^{-1} \cdot cm^{-2}$ at a 1-cm wavelength interval. Along the continuous spectrum of this binary system the wavelengths of the important selected working bands were $\lambda \lambda$  6488, 6563, and 6637. The absolute flux in the H$_\alpha$  emission line above the continuous spectrum in the band $\lambda$ 6563 was considered as a certain physical parameter in view of the interconnection to the variable magnetic field of the donor. Incidentally, the monochromatic light curve reflecting the radiation flux in the band at $\lambda$ 6488 showed that in the second quadrature the binary system is hotter than in the first one. This brightening in the continuum correlates with such excess in the far-ultraviolet \citep{Kondo1994}. However, the dependence of the absolute radiation flux $E(H_\alpha)$ on the orbital phase is more interesting (see Fig.~\ref{fig:1}) and needs more careful consideration.

The upper box in Fig.~\ref{fig:1} shows the variation of the effective magnetic field of the donor as calculated from measurements made on Zeeman spectrograms in 1980-88. This quasi sinusoidal curve shows the maximum value of the magnetic field of negative polarity at the phases around 0.855\,P (the characteristic time of its secondary variations one also can see at these phases). The dipole axis in direction (0.355-0.855)\,P has the 0.145\,P position-angle deviation (one-seventh of the orbital period) from the line joining the centers of stars in the binary system in the direction (0.5-1.0)\,P. In the center of Fig.~\ref{fig:1} the solid line, denoted $E_H$ represents $E(H_\alpha)$ as a function of phase. This dependence was derived from the observation points obtained by moving the phase interval of 0.07\,P in steps of 0.01\,P. The average error per point is less than $\sigma=0.03\cdot 10^{-8} erg/cm^2 \cdot sec$.  Figure~\ref{fig:1} clearly shows that the minimum and maximum values of the radiation flux in the H$_\alpha$ line correspond to the two phases of the extreme values of the magnetic field of the donor. It is seen that the phase variability of the radiation flux of the H$_\alpha$ line has a narrow minimum coinciding in phase with the magnetic field minimum, and the broad maximum of the radiation flux in the H$_\alpha$  line matches with the phase of the maximum value of the magnetic field. The phase intervals of the minimum  $\Delta$\,P$_{min}$=(0.57\,P-0.14\,P) and maximum  $\Delta$\,P$_{max}$= (1.14\,P-0.57\,P) of the radiation flux of the H$_\alpha$ line have the ratio close to 3/4. The radiation flux in the H$_\alpha$ emission line between these extremes of the magnetic field increases 1.3 times, having some constant minimum level in the phase of the minimum of the donor magnetic field.

From this, it can be concluded that if the Roche cavity of the donor is filled, as is generally accepted, it should be assumed that the outermost layers of the donor surface easily lose matter in the direction of the dipole axis of (0.355-0.855)\,P. However, this is observed mainly from the donor surface in the phase region around 0.855\,P, i.e., in phases of the observation of the magnetic field pole facing the gainer. The radiating matter, as a material leaving the donor in the form of a stellar wind along the lines of the magnetic field, deflects hereafter to the gainer and forms a complex structure of accretion flows. And since initially the gas moves perpendicularly to the donor surface, the $H_\alpha$  emission line is generated close to the donor surface and its averaged radial velocities must first have a thermal component, that is, they should not be very large \cite{Skulskii1992-Malkov}. This is confirmed in section~\ref{sec:2_3}, where the radial velocity of the H$_\alpha$ emission line as a whole shows the variability over the orbital period in the range of $\pm45$ km/s.

Another non-trivial, but expected, result was obtained. The solid line middlemost in Fig.~\ref{fig:1} shows, that there is the variable radiation flux of the H$_\alpha$ line at the maximum of the magnetic field, but there is also its explicit excessively outstanding part at the phases of (0.36-0.51)\,P that makes apparent the additional H$_\alpha$ line radiation flux (by 12 percent relative to the smoothed E$_{\rm c}$ curve; see Fig.~\ref{fig:1}, bottom). The phase interval with the width of $\Delta$\,P=0.15\,P can be interpreted as the localization zone of direct collisions of gas flows with the accretion disk at the formation of a hot region on it. This is confirmed by numerous polarimetric Beta Lyrae observations, including those of \cite{Appenzeller1967, Hoffman1998}, and \cite{Lomax2012} where all polarization curves show a pronounced minimum at the phases near 0.47\,P, which is interpreted as a result of enhanced scattering by free electrons near the accretion disk. This is also confirmed by the modeling of the light curve by \cite{Mennickent2013} with the detection at the phases around 0.40\,P of the hotter region located on the accretion disk. However, the absolute spectrophotometry by \cite{Burnashev1991} identified clearly both the center and the phase boundaries (0.43$\pm$0.06)\,P of the hotter region projected onto the accretion disk. 

Hence, the absolute spectrophotometry of the Beta Lyrae system revealed the fact of the apparent synchronous variability over the orbital phase of both the effective magnetic field strength of the donor and the absolute radiation flux in the H$_\alpha$ emission line. This indicates that the donor magnetic field directly reflects the processes of radiation generation in the formed under its influence spatial magnetized gaseous structures at the mass transfer in this binary system. 

\subsection{Magnetic field and the rapid variability of the spectrum in the H$_\alpha$ emission region}
\label{sec:2_2}
The foregoing result is confirmed by a study of \cite{Alekseev1989} of the ultra-fast variability of the Beta Lyrae spectrum in the region of the H$_\alpha$ emission line. Observations of this binary system were conducted for 8 nights (over a 13-day orbital period) in August of 1981, using the 6-m SAO telescope in conjunction with the dissector spectrophotometer. The spectral range investigated, centered on H$_\alpha$ for two dispersions amounting to 0.6 and 0.9 \AA/channel respectively, was 150 \AA~ and 460 \AA. The accumulation time for an individual spectrum was equal to 10.48 and 5.24 seconds, respectively. All the data were obtained with a spectral resolution of 0.6 \AA/channel. Only at the 0.73\,P phase did we make observations using two different options. Based on processing over 700 individual spectra, one should focus on three original results.

\begin{figure}[!t]
	\centerline{\includegraphics[width=0.75\textwidth,clip=]{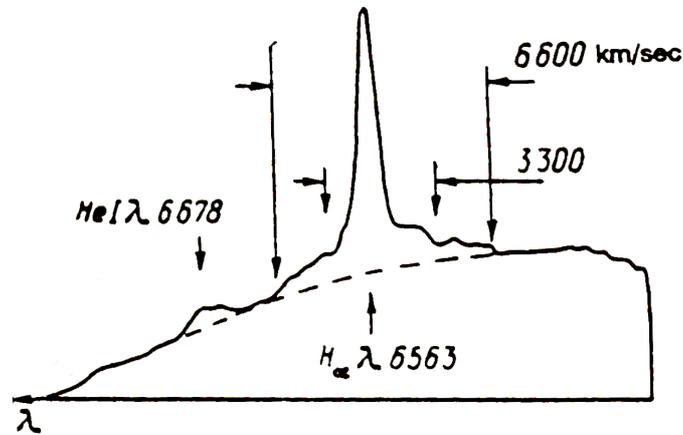}}
	\caption{An average Beta Lyrae spectrum in relative intensities in the 0.73\,P phase (series of spectra with lower spectral resolution): the continuum level under H$_\alpha$ is indicated by the dashed line. }
	\label{fig:2}
\end{figure}

The first important result is the detection of an unusually wide base emission under the strong emission peak in the H$_\alpha$ line that is known to have a height of several continuums and a total width of up to 700 km/s. This broad base emission component has the total width above 6000 km/s, whereas photographic spectral observations, for example by \cite{Batten1973}, show that such a broad component of the H$_\alpha$  emission is only of 1000 km/s. The highest dome-shaped substrate of the broad emission component over 3000 km/s wide rises at its center above the continuum to 25\% in the 0.34\,P phase, up to 30\% in the 0.73\,P phase, and up to 17\% in the 0.87\,P phase, that is in the phases of both quadratures near observation phases of both poles of the magnetic field (see Figure \ref{fig:2} as the fragment of Figure 13 from the article of \cite{Alekseev1989}). On the averaged spectrum with the spectral range of 460 \AA~ at the 0.73\,P phase (88 single spectra), in addition to the H$_\alpha$ line, the He\,I $\lambda$ 6678 emission line was also present, which made it possible to more reliably mark the continuum level. On the violet side of the broad emission component, the narrow absorption with a radial velocity above -2000 km/s is observed in the phase 0.73\,P in the two averaged spectra with the spectral ranges 150 \AA~ and 460 \AA, respectively, and also the absorption at -1000 km/s in the phases 0.34\,P and 0.49\,P. This may indicate high-velocity gas jets from a binary system that are observed in phases near both poles of the magnetic field. 

The high speeds of motion of the radiating atoms, which follow from the fact of the detection of intensive and broad wings of an H$_\alpha$ emission line in quadratures, indicate that the space between the donor and the accretion disk is the zone of the localization of the high-temperature hydrogen medium and of the shock collisions of the speeded-up flows with this disk. The smallest contribution to the continuum in the range of the broad emission component, in particular their distant wings, is observed at phases 0.03\,P, 0.11\,P close to the donor eclipse when the red wing is practically absent (in other words, the high-temperature radiating gas is here almost completely obscured by the accretion disk and only a small amount of this gas moves above this disk in the direction of the gainer, i.e., toward the observer). 

The second important result relates to the rapid variability of spectrum in the H$_\alpha$ line region, which shows significant differences over the orbital phase. The time intervals of this variability range from seconds to tens of minutes. The least emission flux variability is observed in the main minimum at the 0.03\,P phase. There is no appreciable variability in this flux at the 0.88\,P and 0.11\,P phases that are close and symmetric relative to the donor eclipse or the primary minimum of the binary system. In these three phases, the observation region of presumed collisions of the gaseous flows with the disk surrounding the gainer is almost completely eclipsed by this accretion disk (see Figure 1 in \cite{Skulskyy2020}). However, shifting along the positional angle of the observation of the binary system, in particular at the 0.34\,P, 0.42\,P, and 0.49\,P phases, the variability of the radiating flux in the H$_\alpha$  emission line increases substantially: both in the shape and intensity of the central emission peak, as well as within the width of the emission component around this peak. At the 0.34\,P phase, when the space between the components of the binary system is still visible, there is relatively little variability over the entire 15-minute observation. Here, the monotonous variability of the central two-peak profile of the H$_\alpha$ emission line was observed, with fast changes for several one-minute intervals. In the next phase of 0.42\,P, i.e., in the phase of observing the known hot region of the accretion disk, the chaotic variability of the entire emission on 10-second spectra increased very significantly, in both the central peak of the H$_\alpha$ emission line and its broad emission component. However, the most intensive variability of unusual modifications in the entire investigation region of the spectrum near the H$_\alpha$ emission line is observed in the 0.49\,P phase.

\begin{figure}[!t]
	\centerline{\includegraphics[width=0.75\textwidth,clip=]{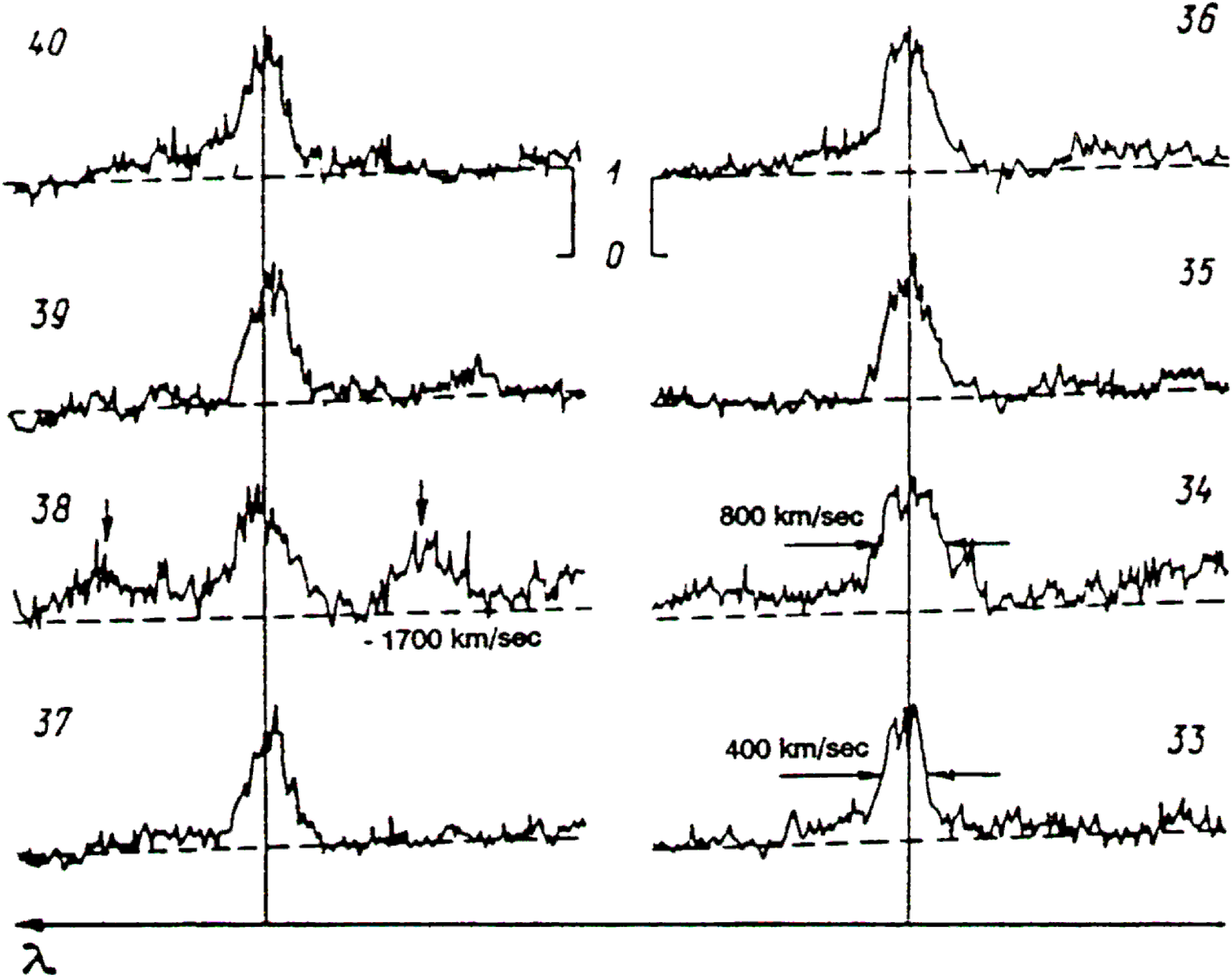}}
	\caption{An example of quantitative estimates on single 10-second H$_\alpha$ line spectra in the 0.49\,P phase: the height of the continuum is shown on spectra 36 and 40; on spectra 33, 34, and 38 there are indicated kinematic estimates of variable details of the spectrum. \citep{Alekseev1989}}
	\label{fig:3}
\end{figure}

The evidence of obtained fast changes in the vicinity of the H$_\alpha$ line in the 0.49\,P phase is shown in Fig.~\ref{fig:3}, which is a copy of Figure 23 from the article of \cite{Alekseev1989}. Time intervals of the continuous variability of different details in this spectrum diapason are observed from seconds to ten minutes. This result is not simple to understand because at the 0.49\,P phase the donor should practically eclipse the known hot region on the accretion disk. At the same time, from the space around this hot region, the gas outflow moves with a great radial velocity of about -1000 km/s, which is recorded in the spectrum by the appearance of absorption lines at phases from 0.34\,P to 0.49\,P. Then, it is possible to assume the formation of the hot pulsating plasma at the upper disk region, not obscured by the donor, as a result of the collision in this region of ionized gas channeled toward the accretion disk due to the specific structure of the donor magnetic field (the magnetic dipole axis is inclined to the orbit plane of this binary system by an angle of 28$\degr$). This may indicate that the moving gas passing along the lines of the magnetic field in the direction of the gainer collides with the accretion disk not so much in the plane of the orbit as the entire height of the disk, forming in front of the disk some scattering structures, which are clearly shown in Figure~\ref{fig:1} as the hotter region on phases (0.39-0.51)\,P. It should be considered that the visible height of the accretion disk in phases of active shock collides may be observed above the donor surface. 

\begin{figure}[!t]
	\centerline{\includegraphics[width=0.75\textwidth,clip=]{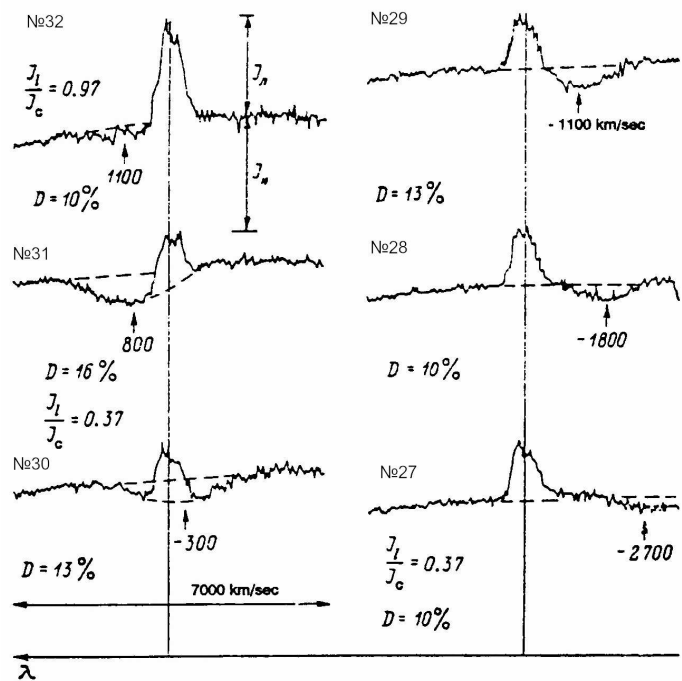}}
	\caption{Detailed illustration and quantitative characteristics of \textquote{traveling absorption} and \textquote{emission flare-up} phenomena (dashed lines indicate the behavior of continuum; the zeros of the continuums coincide with the maxima of the lower spectra). \citep{Alekseev1989}}
	\label{fig:4}
\end{figure}

The third result is similar to the phenomenon of the eruptive nature, which consists of two components: \textquote{emission flare-up} and \textquote{traveling absorption}. It was recorded in the set of 49 spectra at phase 0.81\,P and is shown in Figures 24-28 of \cite{Alekseev1989}. The development of the central part of this phenomenon is illustrated in Fig.~\ref{fig:4}, which is a copy of Figure 28 from \cite{Alekseev1989}. Initially, from the first to the 31st single spectrum, an unusually low ratio of the flux intensity in the central peak of H$_\alpha$ emission line to the emission flux intensity in the continuum near the line of $I_l/I_c$ was observed. This ratio gradually increased from 0.25 to 0.37, but in spectrum 32 reached sharply the value of 0.97, and in 33-49 spectra remained stable at the level of 1.40. Such a sudden increase in the $I_l/I_c$ ratio in limits 31-33 spectra looked like the \textquote{emission flare-up} phenomenon. Such a sharp transformation in the intensity of the H$_\alpha$ emission line was parallel preceded by the phenomenon of \textquote{traveling absorption}, which occurred suddenly within a minute in spectra from 27 to 32. This event began in spectrum 27 with the occurrence of the violet shift absorption at an averaged radial velocity of -2700 km/s from the center of the H$_\alpha$ line. The core of this absorption, over 2000 km/s wide, reached a depth of 16\% in the spectrum 30 and lowered the entire H$_\alpha$ emission line almost under the continuum. The rapid spectral shift of this absorption along the continuum reached the position of +1100 km/s on the red wing of the H$_\alpha$ emission line and disappeared almost entirely in the spectrum 32.

The observed phenomenon has no analog, and it is difficult to unambiguously interpret. Its initial stage was fixed at the moment when the spectral region with the central emission peak of the H$_\alpha$ line had an unusually small contrast with respect to the close continuum ($I_l/I_c = 0.25$ for the first 2-3 minutes from all 8 minutes observations) and was practically imperceptible unlike the formation region of the broad emission component of this line. It is unknown how long this situation lasted before our observations began (it can be noted that it took at least 5 minutes to verify the correctness of the spectral region identification due to the unusualness of the spectrum near H$_\alpha$ line). The emergence, development, and disappearance of this peculiar \textquote{traveling absorption} at its motion along the continuum (within one minute) can be a reflection of a certain mechanism of the matter ejection in the direction of the observer and the reverse movement of this matter of smaller density to the donor surface. This was accompanied by the \textquote{emission flare-up} phenomenon, i.e., the sudden (within half a minute) increase in the $I_l/I_c$ ratio from 0.37 to 1.40 in the central emission peak. However, the characteristic time scale of the variability in the radiation flux of these emissions does not reflect the large-scale motion of gaseous structures commensurate with the Beta Lyrae system. 

The aforementioned phenomenon as the event of the eruptive nature may be related to the spatial configuration of the donor magnetic field. Indeed, this phenomenon was detected at the 0.81\,P phase when the pole of the magnetic field on the donor surface is directed towards the observer (incidentally, \cite{Skulskij1982} observed the maximum of the magnetic field of the donor on the 6-m SAO telescope the same night in August of 1981). This phenomenon is more likely to have local spatial and temporal characteristics close to the donor surface. Such an event may be a reflection of the ejection of a matter directly from the donor surface, on which the location of the magnetic field pole, close to the gainer, is clearly visible in this phase. That is, this event may reflect the motion of magnetically controlled matter, which can be directed outward from the magnetic polar region on the donor surface, as suggested by \cite{Shore1990} for helium stars. This is also just one of the possible spatiotemporal events of different durations recorded by different observers, for example, \cite{Bless1976, Skulskij1980_PAZ}. It is confirmed by \cite{Kondo1994} owing to the observed outburst fixed on the 965 \AA~ band light curve in the same second quadrature. The physical nature of such events may be similar. It can be assumed that the ionized plasma, which is channeled by the magnetic field of the donor from its surface and subsequently collides with the accretion disk, is not laminar, but is accompanied by sometimes significant non-stationary ejections.

\subsection{Magnetic field, dynamics of the H$_\alpha$ emission line and other\\ emission-absorption lines}
\label{sec:2_3}
From foregoing convincing correlations between the variability of the radiation flux in the H$_\alpha$ line and the effective magnetic field strength of the donor, the logical task was to investigate such causal relationships based on the dynamics and structure of the complex $H_\alpha$ emission line. It should be noted that our long-term studies of strong hydrogen and helium emission lines, starting with \cite{Skulskij1972}, have shown that the variability of these lines required obtaining better spectral material. The exact emission-absorption profiles of the H$_\alpha$ lines, represented on a dynamic scale as \textquote{relative intensity versus radial velocity} according to Figure 2 of \cite{Skulskii1992-Malkov}, were obtained using high-dispersion 3 and 6 \AA/mm CCD spectrograms for 20 nights 1985-1990 on the 2.6-m CrAO telescope. This article confirmed the reality of both the rare sharp fluctuations in the H$_\alpha$ flux at intervals of up to 10 minutes and some of the seasonal changes noted in previous studies. More importantly, this article presented the new results of reliable measurements of the Doppler shifts of the structural components of the H$_\alpha$-profile, as well as the factors affecting the profile and dynamics of the H$_\alpha$ emission line as a whole. 

The H$_\alpha$ emission line, as the strongest emission line of the visible spectrum of the Beta Lyrae, exhibits two emission peaks above the continuum and the absorption feature between these emission peaks. The radial velocities of these three variable structural components are traditionally measured. \cite{Skulskii1992-Malkov} measured these line components based on the hypothesis that the H$_\alpha$ emission line as a whole is of a common nature and that the absorption component, which cuts through this emission, originates as self-absorption in this emission line. Under this hypothesis of \cite{Skulskii1992-Malkov}, the radial velocities of the Gaussian center of the total emission as a whole were also measured. The measurements of the radial velocities of structural components of the H$_\alpha$ line showed that all four radial-velocity curves are in one way or another correlated with the effective magnetic field curve of the donor over the orbital period (see Figure 5 of \cite{Skulskii1992-Malkov}). Two types of the V$_r$-curves are identified: 1) those that more clearly correlated with the radial velocity curves of the components of the binary system, i.e., related in space to the line of their centers, which passes through the direction of the phases of (0.5-1.0)\,P; and 2) those that more clearly correlated with extrema of the effective magnetic field of the donor, i.e., with the spatial direction of the dipole axis of its magnetic field of (0.355-0.855)\,P (see Figure~\ref{fig:1}). The first type of curve is the radial velocity curve of the absorption feature, which shows negative radial velocities at all phases reflecting the gas flows motion. The radial velocity curve for the Gaussian center of the H$_\alpha$ emission is definitely of the other type: both its maxima clearly match extrema of the effective magnetic field strength of the donor. Hence, such an approach to the measurement of Doppler shifts of the characteristic features of the H$_\alpha$ emission-absorption line as a function of the orbital phase has revealed two major factors forming its shape and dynamics. The more important one reflects the Doppler phase shifts of the H$_\alpha$ emission profile as a whole. It follows that the generation of the radiation flux in the H$_\alpha$ emission line and the formation of the emission-absorption profile of this line can be formed under the certain influence of the donor magnetosphere. 

\begin{figure}[!t]
	\centerline{\includegraphics[width=0.75\textwidth,clip=]{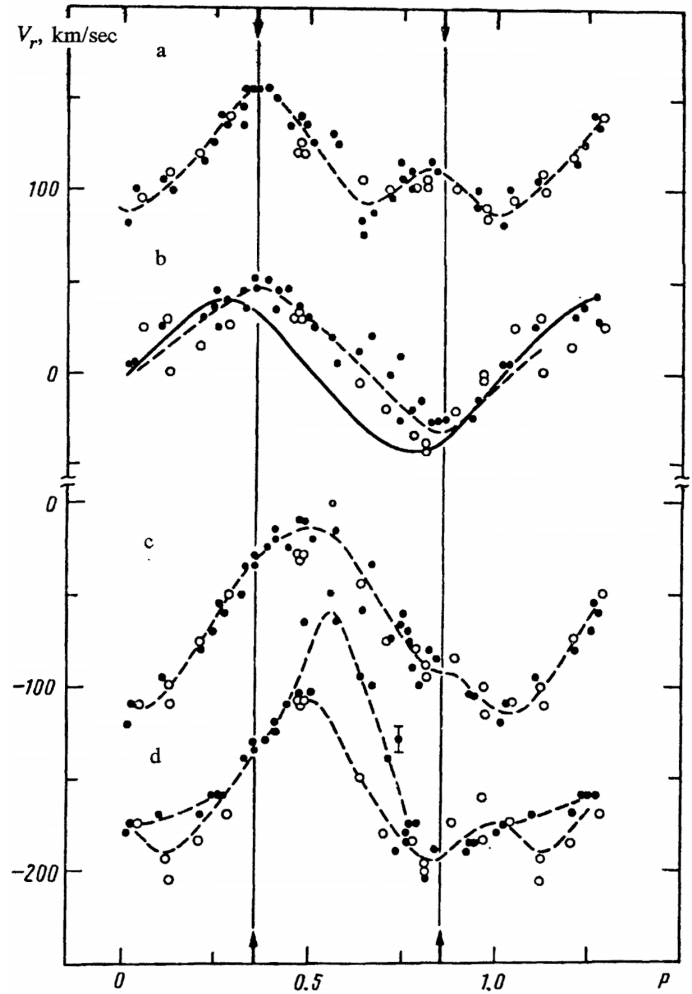}}
	\caption{Radial-velocity curves of structural components of the H$_\alpha$ emission profile based on CCD observations in 1990-91 (dots) and 1985-89 (circles) for: a) the redshifted peak; b) the center of the Gaussian profile H$_\alpha$ emission (the solid curve is the V$_r$ curve for the gainer); c) the absorption feature cutting through the emission; d) the blueshifted peak (the mean error $\approx2\sigma$ of an individual measurement is shown by a vertical bar). The dashed curves were drawn by hand. The zero points on the vertical axis coincide with the zero velocity of the binary system's center of mass. Two extrema of the curve of the effective magnetic field strength are marked by arrows and vertical lines at the phases 0.355\,P and 0.855\,P. \citep{Skulskij1993b}}
	\label{fig:5}
\end{figure}

These results changed the understanding of the formation of gas structures and required confirmation. Therefore, in the following year, 1991, the intensive CCD observations of the Beta Lyrae in the red spectral range were extended. They included the simultaneous detailed study of general regularities in the phase variability of the parameters of emission-absorption lines $H_\alpha$, He\,I  $\lambda \lambda$ 6678, 7065 and Si\,II $\lambda \lambda$ 6347, 6371. The investigation of circumstellar structures and mass transfer in the presence of the donor magnetic field, which was based on a study of the dynamical and energy characteristics of these lines, is given by \cite{Skulskij1993b}. Figure~\ref{fig:5}, which is presented here as Figure 3 from \cite{Skulskij1993b}, shows quite sufficient statistical reliability of the four radial velocity curves of the components of the H$_\alpha$ emission line, which were obtained with the incorporation of the data from \cite{Skulskii1992-Malkov}. Figure~\ref{fig:5} illustrates also the phase coordination of these radial velocity curves with the curve of the effective magnetic field of the donor (see Fig.\ref{fig:1}) that needs careful consideration.

First of all, in Fig.~\ref{fig:5}, the measurements of the Doppler shifts of the H$_\alpha$ profile components are presented separately for 1985-89 and 1990-91. In the later period, clear differences in the radial velocity curves of the blueshifted (or violet) peak are revealed, indicating certain long-term changes in the gas emitting structures directed mainly to the observer. These changes are observed in phases (0.0-0.15)\,P, but especially in phases from 0.50\,P to 0.75\,P of the second quadrature (see Fig.~\ref{fig:5}, bottom). The shape of this curve is clearly changed in 1990-91 at a virtually invariable shape of the radial velocity curve of the absorption component that points out the independence of the moving plasma radiated in the violet emission peak. Indeed, the minimum value of negative radial velocity fell from -100 km/s to -50 km/s diminishing accordingly the distance in radial velocities between the violet and red emission peaks, and the width of the H$_\alpha$ emission as a whole. Moreover, the violet emission peak shifted to the phases of (0.55-0.60)\,P, reflecting a certain direction perpendicular to the direction of the axis of the donor magnetic field of (0.35-0.85)\,P. It is important that during all the years at the 0.85\,P phase, which corresponds to the phases of the maximum magnitude of the donor magnetic field and the visibility of the magnetic field (see also Fig.~\ref{fig:1}), this violet peak shows the maximum value of the negative radial velocity of -200 km/s. It could be interpreted as the additional matter outflow from the region of the magnetic field pole of the donor surface, facing the gainer, along the direction of the magnetic field axis of (0.35-0.85)\,P.
 
A clear correlation is observed between the radial velocity curves of the star-components of the binary system and the radial velocity curve of the absorption on $H_\alpha$ emission profile. This almost symmetric V$_r$-curve with the shape of a dome around the 0.55\,P phase shows here the minimum radial velocity of -15 km/s of matter in the direction from the binary system. This is close to the thermal velocity from the surface of the bright donor that reaches to its Roche cavity. At the phases of both quadratures, the radial velocity of the outflowing matter reaches  -80 km/s. The radial velocity curve becomes slightly asymmetric at the 0.85\,P phase of the magnetic field maximum and shows the maximum of negative velocities near -115 km/s (close to the parabolic velocity of the moving gas) at the 0.05\,P phase when the stellar wind is converted into the fast flow up to the backside of the accretion disk. Both lower V$_r$-curves in Fig.~\ref{fig:5} reflect also the spatial structure of the binary system: they are almost symmetrical about the line of centers of its components, which passes through the phases of 0.5\,P and 1.0\,P. However, the two extrema of these curves reproduce the direction of the phases close to (0.6-1.1)\,P, i.e., the spatial line that is turned on a quarter of the orbital period relative to the dipole axis of the donor magnetic field. This factor should also be taken into account in the study of moving magnetized accretional structures. 

The radial velocity curves for the Gaussian center of H$_\alpha$ emission for 1985-89 and 1990-91 are definitely of the other type: both maxima of this sine curve clearly correspond to two maxima of the curve of the effective magnetic field strength of the donor (see also Fig.~\ref{fig:1}), indicating that these two curves are physically related. The radial velocity curve of the H$_\alpha$ emission center is clearly shifted by 0.1\,P from the radial velocity curve of the gainer (see Fig.~\ref{fig:5}). The radial velocity curve for the more intense red emission peak has also maxima at those phases. It should be noted that the low local maximum in the V$_r$-curve of this red peak at the 0.85\,P phase becomes dominant if one constructs the dependence  $\Delta V_r=f(P)$, where $\Delta V_r$ is the difference between the radial velocities of the long-wavelength peak and the center of H$_\alpha$ emission (see \cite{Skulskii1992-Malkov}). Both upper V$_r$-curves in Fig.~\ref{fig:5} reflect the phase variability of the donor magnetic field, whose dipole axis is directed along the line of the (0.355-0.855)\,P phases, i. e., the general emission as a whole and the red peak as components of the H$_\alpha$ emission-absorption line are essentially formed in the agreement with the spatial configuration of the donor magnetic field.

The foregoing should be considered in conjunction with the careful work of \cite{Sahade1959}. The Beta Lyrae spectrum in $\lambda \lambda$ 3680-4580 was studied on the basis of spectrograms obtained on the Mount Wilson 100-inch reflector. The radial velocities of red emission peaks have been measured in lines of He\,I $\lambda \lambda$ 3888, 4472 and H$_\gamma$ line. They are plotted in Figures 18, 19 and 20 of  \cite{Sahade1959}. These radial velocity curves are in good agreement with the radial velocity curve of the red peak in the $H_\alpha$ line that is shown in Fig.~\ref{fig:5}. This is especially true of the two maxima in the radial velocity curve of the strong red peak in the He\,I $\lambda$ 3888 line. They clearly correspond to the phases of the two extrema on the curve of the effective magnetic field strength of the donor and phases of the visibility centers of two magnetic poles on the donor surface. Such coincidence factors are logical. Their interconnection seems indisputable.

\begin{figure}[!t]
	\centerline{\includegraphics[width=0.75\textwidth,clip=]{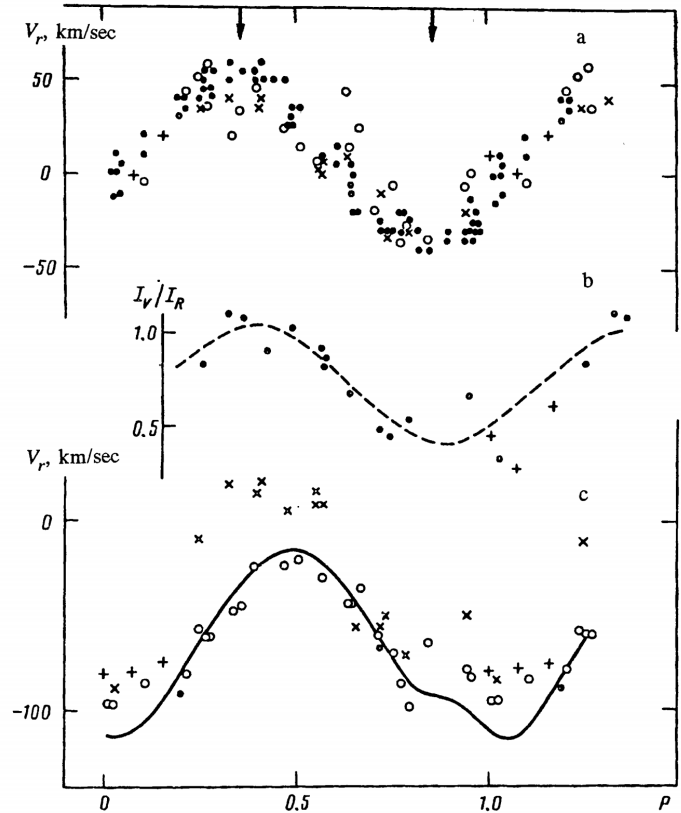}}
	\caption{Radial-velocity curves of a) the center of the Gaussian emission profile in the Si\,II $\lambda \lambda$ 6347, 6371 doublet lines (dots), the He\,I $\lambda$ 6678 line (circles), and the He\,I $\lambda $ 7065 line ($\times$), supplemented by the 1992 observations (crosses); c) absorption features in the same helium lines (1992 observations in the He\,I $\lambda$ 7065 line are represented by crosses and the He\,I $\lambda$ 5875 line are represented by dots); the solid curve here corresponds to the V$_r$-curve of the absorption feature in the H$_\alpha$ line from Fig.~\ref{fig:5}; b) the $I_V/I_R=f(P)$ curve for emission in the He\,I $\lambda$ 7065 line, supplemented by the 1992 observations (crosses). Two extrema of the curve of the effective magnetic field are indicated (at the top) by arrows at the phases 0.355\,P and 0.855\,P. \citep{Skulskij1993b}.}
	\label{fig:6}
\end{figure}

Similar conclusions can be drawn from the consideration of Fig.~\ref{fig:6}, reproduced from Figure 4 in \cite{Skulskij1993b}, as to analogous components in the emission-absorption lines of He\,I $\lambda \lambda$ 6678, 7065 and Si\,II $\lambda \lambda$ 6347, 6371 (mainly based on CCD observations in 1991). Indeed, these V$_r$-curves have extremal values near certain phases of the orbital period: 0.0\,P and 0.5\,P, as well as 0.35\,P and 0.85\,P, i. e., are related to the geometry of the eclipsing binary system and the dipole structure of the donor magnetic field. The orbitally modulated V$_r$-curves of these absorption features have practically sinusoidal symmetry with regard to the axis of the gravity centers of both components. Only such V$_r$-curve for the He\,I $\lambda$ 7065 line is shifted somewhat to the left, to the 0.35\,P phase, which coincides with the phase of the observation of the donor magnetic pole. Reaching a positive radial velocity of more than +15 km/s here, this V$_r$-curve remains positive within (0.3-0.6)\,P. This new fact indicates that the outflows of the radiating plasma with the thermal velocity from the donor surface along the donor axis from the observer is observed here, i.e., beginning in the direction of the 0.85\,P phase of the second magnetic pole and with the next deviation to the gainer. Such matter outflows showed also in the triplet helium line and are probably raised above a denser stream of mainly hydrogen plasma, but reflected primarily in the absorption feature of the H$_\alpha$ line. The influence of the magnetic field at the 0.85\,P phase is clearly visible both on the shape of all absorption V$_r$-curves in Fig.~\ref{fig:6} and in the absorption feature of the H$_\alpha$ emission line in Fig.~\ref{fig:5}.
 
Comparing Fig.~\ref{fig:5} and \ref{fig:6}, one should note good agreement between the average V$_r$-curves for the emission centers of the H$_\alpha$ line and lines He\,I $\lambda \lambda$ 6678, 7065 and Si\,II $\lambda \lambda$ 6347, 6371. The He\,I $\lambda$ 6678 line exhibits the greatest uncertainty when measuring radial velocities due to the substantial asymmetry of the emission profile. However, the radial velocities of the powerful He\,I $\lambda$ 7065 line, whose double-peaked emission, like that of H$_\alpha$ line, is observed considerably above the continuum over the orbital period, are measured reliably. Together with the radial velocities of the emission centers of the red silicon doublet lines (their values were taken from \cite{Skulskij1992}) they mimic well the average V$_r$-curve of the center of H$_\alpha$ emission as a whole. Thus, two sinusoidal extrema of these dependences $V_r=f(P)$ coincide (see Fig.~\ref{fig:5}b and Fig.~\ref{fig:6}a) with the phases of passing through the meridian of both poles of the magnetic field on the donor surface (see also Figure 1 in \cite{Skulskyy2020}). In general, it reasonably suggests that the emission and absorption components of these studied lines are produced with almost the same dynamics in gaseous structures between components of this binary system, correlating clearly in orbital phases with the variability of the donor magnetic field. 

In view of the above, it is worth noting the thorough work of \cite{Harmanec1996}, which, in particular, presents the results of the processing of the electronic spectra obtained in 1992 and 1994 on the reflectors of Ondrejov and Dominion Astrophysical Observatories in the red spectral range 6100-6700\AA. Using also the data of our publications \citep{Skulskij1972, Skulskij1993b, Skulskii1992-Malkov}, they presented statistically better radial velocity curves of the H$_\alpha$ emission line for all four of its components (see Figures 8 and 9 in \cite{Harmanec1996}). Their behavior of $V_r=f(P)$ or \textquote{RV curves}, as expected, are in full agreement with the corresponding radial velocity curves in Fig.~\ref{fig:5} and \ref{fig:6} of this paper. Their \textquote{RV curves of the $H_\alpha$ and He\,I $\lambda$ 6678 absorption cores versus orbital phase} show also the minimum -15 km/s and maximum -113 km/s of the negative velocities near the 0.55\,P and 0.05\,P phases, respectively. The extrema of their RV curves for both emission peaks of the H$_\alpha$ line clearly correspond to the phases of the observation of the poles of the donor magnetic field or the direction (0.355-0.855)\,P. Incidentally, the same maximum negative velocity of -200 km/s was shown for the violet emission peak (or \textquote{V-peak}) in the 0.855\,P phase. These investigators indicate that \textquote{the H$_\alpha$ emission as whole moves almost in anti-phase} to the donor, such that for the radial velocity curve of this emission \textquote{the RV minimum is 0.853\,P (instead of 0.75\,P for the donor)}. This coincides with the maximum of the donor magnetic field at the 0.855\,P phase (see Fig.~\ref{fig:1}), as it follows from the solution of the magnetic field variability curve by \cite{Burnashev1991}. And this corresponds to the phase meridian of the passage of the magnetic field pole on the donor surface. It should be noted that in interpreting the Beta Lyrae system \cite{Harmanec1996} do not use such a clear match in these results and the studies of the donor magnetic field as a whole. They concluded: \textquote{the bulk of the H$_\alpha$ and He\,I $\lambda$ 6678 emission seems to originate in jets of material perpendicular to the orbital plane of the binary}. Here these jets are associated mainly with the gainer and also \textquote{probably emanate from the hot spot in the disk, i.e. the region of the interaction of the gas stream flowing from the Roche-lobe filling} of the donor. On the contrary, \cite{Ignace2018} are hopeful that the magnetic field \textquote{could prove relevant}. Based on the study and simulating of H$_\alpha$ line profile variations in the Beta Lyrae spectrum, they have shown that \textquote{a circumbinary envelope, a hot spot on the accretion disk, or accretion stream} non-satisfactory explain the observation data. They even presume: \textquote{evaluating the detailed radiative transfer for a model involving both a disk and a jet is unlikely to help}.

An important conclusion that the emission as a whole is produced mainly in structures that are affected by the magnetic field, that is, they are largely generated near the donor surface, was given by \cite{Skulskij1993a, Skulskij1993c}. This was also based on variations of V$_r$-curves of the emission center Si\,II $\lambda \lambda$ 6347, 6371 lines in 1990-92. There are two irrefutable facts to such conclusion: a shift near 0.1\,P in the phase of the V$_r$-curve of this emission as a whole, and a shift of about +10 km/s of the center of the radial velocity of this V$_r$-curve relative to the radial velocity of the mass center of the binary system. These factors should be considered based on Fig.~\ref{fig:5} and \ref{fig:6}. First, a clear difference of 10 km/s between the radial velocity of the mass center of the binary system and the velocity of the emission center of all V$_r$-curves implies a constant outflow of matter from the binary system. Then it is necessary to associate this outflow of matter not with the gainer but with the bright donor, which is \textquote{constantly expanding} toward its inner Roche cavity. The matter drains essentially and along the magnetic field lines from the donor surface near the magnetic pole facing the gainer, i.e., near the 0.85\,P phase (see Figure 1 of \cite{Skulskyy2020}). The maximum negative shift of radial velocities of -45 km/s in the V$_r$-curves of the emission centers of all the investigated lines is detected at just those phases (i. e., the motion from the surface of the donor toward the observer). The maximum positive shift of these radial velocities of +45 km/s is recorded in the vicinity of the opposite magnetic field pole around the 0.35\,P phases, i.e., the emitting gas moves from the observer (probably along the magnetic field lines), but in the same direction as its the motion registered around the 0.85\,P phases. That confirms that gas outflows are produced mainly on the donor surface near the donor's magnetic pole, which is close to the massive gainer and observes near the 0.85\,P phase. At the same time, the V$_r$-curves of the emission centers of all the investigated lines are shifted in phase by 0.1\,P relative to the V$_r$-curve of the gainer, and the sinusoidal maxima of these V$_r$-curves coincide with such maxima on the curve of the donor magnetic field (see also Fig.~\ref{fig:1}). This is strong evidence of the physical relationship between the regions of spatial formation of the total emission and the structure of the donor magnetic field. One can also consider as reasonable the picture, in which the magnetized plasma, moving from the donor surface mainly in the 0.85\,P phases and deflecting afterward to the gainer, forms a system of gas flows directed toward the accretion disk (this is also seen from the radial velocity curve of the absorption component of the H$_\alpha$ emission line).

Hence, it should be stressed that the variability of the complex profiles of the emission-absorption lines over the orbital period, in terms of their dynamic characteristics, is essentially synchronous with such variability of the magnetic field of the donor. The reflection of accretion flows in the spectrum of this interacting binary system during the mass transfer from the donor to the gainer is largely determined by the existing spatial structure of the magnetic field of the donor.

\subsection{Magnetic field and investigation of energy characteristics of spectral lines formed near and in the atmosphere of the donor}
\label{sec:2_4}
From the above it follows that the donor magnetic field significantly influences the overall picture of the localization and formation of developed gaseous structures. A more active region of the outflow of matter from the donor surface is the region of ​​the magnetic pole facing the gainer. A mass-losing donor with the decentered magnetic dipole has a deformed surface, reaching its Roche cavity. This stimulated a parallel study of the phase variability in the energy characteristics of different lines, which are formed in both the conditions of this peculiar atmosphere and the neighborhood of the donor surface. This applies, above all, to spectral observations made in the red spectral region in 1985-1992 using a CCD detector on the 2.6-m CrAO telescope and, in particular, a comprehensive study of all components of the complex structure of the Si\,II $\lambda \lambda$ 6347, 6371 lines \citep{Skulskij1993a}. The phase variability of the equivalent width $W_\lambda=f(P)$ and $I_l/I_c=f(P)$, i.e., the emission intensity ratio in the line to the continuum, in the emission component of Si\,II $\lambda \lambda$ 6347, 6371 lines shows that there is a well-defined modulation of these Si\,II emission curves by the donor’s magnetic field.  

The phase dependence of the intensity ratio of the violet to red emission peaks $I_v/I_r=f(P)$ in the He\,I $\lambda$ 7065 line (see Fig.~\ref{fig:6}) was also investigated by \cite{Skulskij1993b}. The emission of the He\,I $\lambda$ 7065 line is interesting because its emission peaks are comparable in intensity, and this is the only line where its violet peak even slightly exceeds the red peak over a considerable phase range close to the 0.35\,P phase, that is in the range of phase visibility of the donor's magnetic field pole. The minimum ratio of their intensity (i.e., when the red peak dominates) coincides with phases of visibility of the opposite pole of the donor facing the gainer, i.e., of 0.85\,P. Thus, the dependence of $I_v/I_r=f(P)$ is also clearly modulated by the spatial structure of the donor's magnetic field. It also means that the emission in both the strong He\,I $\lambda$ 7065 line and Si\,II $\lambda \lambda$ 6347, 6371 lines, with their relatively not high emission, is produced largely in some medium in the immediate vicinity of the donor. In addition, \cite{Skulskij1993b} also investigated the phase dependence of the ratio $\Delta W_\lambda /W_\lambda$, i.e., of the equivalent width of the absorption component in a range of the total emission to the total emission under the Gaussian profile in H$_\alpha$ and He\,I $\lambda$ 7065 lines, which can explain the variability of the self-absorption in this emission. For both lines, this parameter has maxima in phases that coincide with the phases of the two magnetic field poles visibility. In this article, phase variations of the equivalent width of the He\,I $\lambda \lambda$ 3867, 4120 absorption lines of the donor's atmosphere and the total emission in the He\,I $\lambda$ 7065 as lines of one helium triplet were also studied. Phase cyclic changes in the equivalent width inherent in the He\,I $\lambda \lambda$ 3867, 4120 absorption lines of the helium triplet, mainly being synchronized with such $W_\lambda$-curve for the He\,I $\lambda$ 7065 emission line, are found, that is, they have much in common, which is related to the structure of the donor magnetic field. 

The investigation of complex Si\,II $\lambda \lambda$ 6347, 6371 lines also revealed the apparent correlation between the phase variability of the equivalent width of Si\,II-emissions and such $W_\lambda$-curves of absorption lines of this doublet in the donor atmosphere \citep{Skulskij1993a}. In addition to the effects of the orbital modulation and eclipses, there is a well-defined modulation of these curves by the magnetic field. One also concluded that the formation of emissions in these silicon lines should be localized in the immediate vicinity of the donor surface. This finding was supported by a new fact: the deep minimum was found on the W -curves of these Si\,II emissions in a narrow phase range of 0.02\,P duration before the main eclipse of the donor at the phase of 0.96\,P in 1991 and 0.93\,P in 1992 \citep{Skulskij1993a, Skulskij1993c}. At these phases, the satellite disk, as the outer part of the accretion disk, is projected onto the magnetic pole region on the donor surface facing the gainer, significantly eclipsing the source of the emission in the red Si\,II doublet (see Figure 1 in \cite{Skulskyy2020}). This can also be interpreted as an eclipse of the hot region of the donor surface or near this surface, i. e., in the direction of the donor magnetic field pole close to the gainer. 

The unusual behavior of the equivalent widths of both absorption lines in the red Si\,II doublet of the donor's atmosphere led to a detailed study of the phase variations of more than 100 absorption lines of this atmosphere in the blue spectral range, published by \cite{Skulskij1971} in the ten main phases of the orbital period. As it turned out, the equivalent widths of the lines of the magnetized and simultaneously outflowing atmosphere of the donor exhibit a special kind of cyclic variations over the orbital period \citep{Skulskij1993a}. The $W_\lambda$-curves of these lines are modulated to varying degrees both by gravitational and magnetic fields. Dozens of the $W_\lambda$-curves look like the result of the superposition of harmonic oscillations with different amplitudes and frequencies that are multiples of the orbital frequency. We tried to classify the $W_\lambda$-curves in terms of their external form, taking into account the patterns of splitting in the magnetic field and the value of the Lande factors, the total angular momentum of the atoms and the relative intensity of lines in their multiplets, the degree of excitation and ionization. No definite regularities have been discerned, but virtually all absorption lines can be grouped into three types. The first type includes the lines whose $W_\lambda$-curves are more definitely subject to orbital modulation. These are more excited or resonance lines, e. g., more intense lines of dominant FeII and TiII multiplets, for which the maxima of their equivalent widths occur, as a rule, at phases around 0.0\,P and 0.5\,P, i.e., related in space to the line on star-components. It may seem that this group of lines originates in the upper layers of the donor atmosphere, which is extended to its Roche cavity and has a somewhat elongated surface, especially in the gainer direction. The equivalent widths of the second group lines are more clearly correlated with the phase variability of the donor magnetic field, i.e., with the direction of the dipole axis of its magnetic field passing through the 0.35\,P and 0.85\,P phases. This group includes, as a rule, not very strong absorption lines and lines of a higher degree of excitation and ionization, i.e., lines originating in the relatively deeper layers of the atmosphere. The third, intermediate, group of lines demonstrates the variations in their equivalent widths as a reflection of possible simultaneous actions of the gravitational and magnetic fields.

 It is most likely that the demonstrated diagrams of phase variations of the equivalent width and intensity of the different lines of visual spectrum reflect real changes in the physical conditions, both with the depth of the donor atmosphere and above the atmosphere level. However, this occurs both under the conditions of the specific spacial structure of the donor magnetic field and its deformed surface close to its Roche cavity. This can be seen, for example, from gradual changes in the profiles of lines of the Balmer series. The first members of the series, as shown in the previous sections, are formed above the surface of the donor and further in the moving gaseous structures. They exhibit strong emission components that gradually decrease in intensity and fall under the continuous spectrum close to H7. The high terms of the Balmer series, which are undistorted by emission, according to Figure 6 in \cite{Skulskij1993a}, show the notable systematic change of the $W_\lambda$ -curves from line to line over the orbital period from H9 to H21. In particular, in $W_\lambda$ -curves of lines H11 - H16 there are here definite narrow local maxima. At the same time, equivalent widths of lines H17 - H20 show here very deep local minima; moreover, during the orbital turning of the donor from the 0.5\,P phase to the phase of 0.85\,P, the equivalent widths of lines H17 - H19 decreased to one-third of their initial widths. All this going out that the behavior of the spectral lines of the Balmer series, from its high terms up to the H$_\alpha$ line, reflects a certain change in physical conditions in the donor atmosphere and above its surface. These data also indicate the most pronounced stratification of these conditions in the phase range (0.85 $\pm$  0.15)\,P at the passing of the observer above the donor surface magnetic pole. It could be indicated that similar phase diagrams of variations of the equivalent widths for these hydrogen lines are shown in \cite{Bahyl1986}, where there also were noted coordinated phase variations of the equivalent widths for some spectral lines of metals and neutral helium in the donor's atmosphere with phase variations of the donor magnetic field. Hence, this can be seen as a confirmation of our understanding and interpreting mass transfer processes in the Beta Lyrae system based on the concept of the formation of magnetized accretion structures at the presence of a certain spatial configuration of the donor magnetic field.

\subsection{Magnetic field and the identification of the donor magnetic pole facing the gainer}
\label{sec:2_5}
In view of the above, the next important task was, on the basis of different observations, to attempt to directly identify on the donor surface the range phase visibility of the magnetic pole facing the gainer. 

It should draw attention in this regard the accurate infrared photometry of the Beta Lyrae system in 5 bands from J (1.2 $\mu$m), H (1.6 $\mu$m), K (2.2 $\mu$m) to L (3.5 $\mu$m) and M (4.6 $\mu$m) that was carried out by \cite{Zeilik1982} in 1977-1982 on the 1.3-m telescope at the Kitt Peak National Observatory. In these observational data, several points that need more attention are apparent. The general shape of these light curves is similar to the visual light curves. The data, collected in Figures 1-5 pursuant to the J, H, K, L, and M filters as for observations in different years (they are demonstrated as magnitude differences of ($\beta$ Lyr - $\alpha$ Lyr) at statistical errors per point that are typically less than the symbol size), have a good convergence. As expected (according to the previous article of \cite{Jameson1976}, where observations were also made using $\alpha$ Lyrae as a standard star), the depth of the primary eclipse decreases relative to the depth of the secondary eclipse at longer wavelengths and the magnitude differences (in the mag) are such: 0.28 at J, 0.24 at H, 0.12 at K, -0.06 at L, and 0.16 at M bands. Equality of eclipses depths occurs at ~ 3 $\mu$m. Previous observations of \cite{Jameson1976} also clearly showed a greater secondary eclipses depth in the close bands at 3.6 $\mu$m, 4.8 $\mu$m, and especially at 8.6 $\mu$m (the last band is absent in \cite{Zeilik1982}). The difference between the maxima on the light curve in the quadratures is particularly noticeable at 8.6 $\mu$m, where the asymmetry of the light curve is 0.4 mag, with the maximum of the light curve near the 0.35\,P phase, i.e., in the phases visibility of the magnetic pole on the donor surface in the primary quadrature. \cite{Jameson1976} noted that at 3.6 $\mu$m, 4.8 $\mu$m, and 8.6 $\mu$m the infrared light curves have also some fine structure in the phases of the secondary quadrature,which is important in the analysis of the data of \cite{Zeilik1982}.
 
\begin{figure}[!t]
	\centerline{\includegraphics[width=0.75\textwidth,clip=]{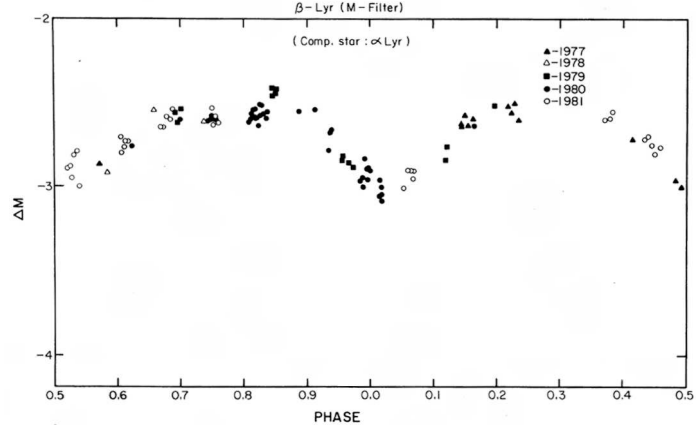}}
	\caption{Raw data as magnitude differences ($\beta$ Lyr - $\alpha$ Lyr) for $\beta$ Lyr at M band (4.6$\mu$m). Dates are, 1977 (closed triangles), 1978 (open triangles), 1979 (closed square), 1980 (closed circle) and 1981 (open circle). Statistical errors per point are typically less than the symbol size  \citep{Zeilik1982}. }
	\label{fig:7}
\end{figure}

Our review of the observational data in both articles cites that, in the secondary quadrature near the 0.85\,P phase, on the four light curves in the range of 3.5-4.8 $\mu$m one can see clear local segments of the radiation increase. They permit in this phase the direct identification on the donor surface of the magnetic field pole region close to the gainer. This draws up primarily from the more accurate M band data in Fig.~\ref{fig:7}, which is introduced as Figure 5 from the article of \cite{Zeilik1982}. This light curve has the secondary eclipse depth of -2.92 mag in phase 0.5\,P and, reaching a value of -2.6 mag in phase 0.68\,P, remains practically invariable until phase 0.80\,P. From phase 0.81\,P the light curve increases abruptly by 0.20 mag to phase 0.85\,P, reaching here the maximum value near of -2.4 mag, and then decreases to the previous value of -2.6 mag in phase near 0.90\,P and to -3.08 mag in phase 0.0\,P of the primary eclipse. According to Table~1 and Figure~1 of \cite{Jameson1976}, the light curve at the band of 4.8 $\mu$m for averaged observations in the 0.81\,P and 0.88\,P phases also show the identical increase of 0.2 mag, respectively, from -2.65 mag to -2.45 mag, followed by a rapid drop on the light curve to the 0.95\,P phase (these observations were not performed often enough). As one can see from Fig.~\ref{fig:7}, a sharp and clear radiation increase in the light curve at the M band demonstrates the effective phase width of visibility of this hot region close to 0.1\,P. The maximum in this light curve coincides with the maximum of the magnetic field curve at the 0.855\,P phase \citep{Burnashev1991}, which corresponds to the phase of the meridional passage of the magnetic pole on the donor surface (see also Fig.~\ref{fig:1}). 

Hence, the light curves at M band (4.6 $\mu$m) from \cite{Zeilik1982} and at the band of 4.8 $\mu$m from \cite{Jameson1976} can be quite convincing evidence of the spatial identification on the donor surface of the magnetic field pole facing the gainer. The result of this kind is understandable because at these wavelengths the optical depths are smaller, and this allows one to see the radiating matter above the donor surface deeper, where the effective temperatures are higher. This identification is confirmed from the point of view measuring the Zeeman splittings in the Si\,II $\lambda \lambda$ 6347, 6371 lines according to observations with the CCD detector on the 2.6-m CrAO telescope \citep{Skulsky1993}. The obtained curve of effective magnetic field strength in the Si\,II lines changes with the orbital phase synchronously with the photographic curve of the magnetic field shown in Fig.~\ref{fig:1}. However, on these curves in Si\,II lines there are two regions in which one sees the locally changing polarity of the magnetic field with a width of approximately 0.08\,P (see Figures 5 and 6 in \cite{Skulskyy2020}). Their centers are located exactly in the phases 0.355\,P and 0.855\,P of both poles of the magnetic field on the donor surface. This coincidence can be considered as the independent identification of the magnetic poles on the donor surface. On the other hand, the hot region on the donor surface in the phases around the pole of the magnetic field facing the donor was confirmed by the independent researches of the spectrum near the Si\,II $\lambda \lambda$ 6347, 6371 lines by \cite{Skulskij1993a, Skulskij1993c}. First, this region on the donor surface is identified as the clear-cut of its deep eclipse when the satellite disk, as the outer part of the accretion disk, is passing above the donor surface (see Figure 4b in \cite{Skulskij1993c}). Second, these CCD observations of 1991 and 1992 showed that the phase variations of both the equivalent widths of these absorption lines of the donor atmosphere and the emission component of these lines practically identically reflect the phase variations of the donor the magnetic field, confirming that the regions of formation of all these lines are close to each other.
 
One can summarize those methodically independent observations and investigations of \cite{Zeilik1982, Jameson1976, Skulsky1993}, and \cite{Skulskij1993a, Skulskij1993c}. They admit evidence of the phase boundaries of orbital identification of the magnetic pole facing the gainer on the donor surface and argue that the emissions of the Si\,II $\lambda \lambda$ 6347, 6371 lines are generated actually fairly close to the donor surface, mainly around the location region of this magnetic pole. Moreover, it should be supposed that a hot region centered at the phase of 0.80\,P, which was interpreted in \cite{Mennickent2013} as detected on the accretion disk (see Figure 1 in \cite{Skulskyy2020}), may, in fact, be caused by additional radiation in these phases directly from the magnetic-polar region on the donor surface. This is one of the consequences of the spatial identification on the donor surface the magnetic pole facing the gainer. 

It is now pertinent to note that the identification at the donor surface of additional radiation in the magnetic pole region, as well as of the phase boundaries of this radiation, is important for understanding the nature of the developed gaseous structures in the Beta Lyrae system. In this binary system, as in helium stars with the magnetically controlled matter, magnetized plasma can be directed outwards, forming matter outflows from the magnetic polar region. Indeed, \cite{Shore1990} proposed a known model for helium stars, the essence of which is that \textquote{circumstellar plasma is trapped in the stellar magnetosphere near the magnetic equator or is channeled to form jetlike outflows from the magnetic polar regions}. With regard to the first statement, it is worth mentioning the phenomenon of depression at $\lambda$ 5200\,{\AA} in the continuum of the Beta Lyrae system, which was discovered and studied as an independent argument for the reality of the magnetic field \citep{Burnashev1986}. Both maxima of the equivalent width of this depression are observed in the same phases as the zones of both poles of the magnetic field on the donor surface. At the same time, both minima of the equivalent width of this depression are clearly displaced by a quarter of the orbital period, indicating that the direction perpendicular to the axis of the magnetic field may be important.

\begin{figure}[!t]
	\centerline{\includegraphics[width=0.85\textwidth,clip=]{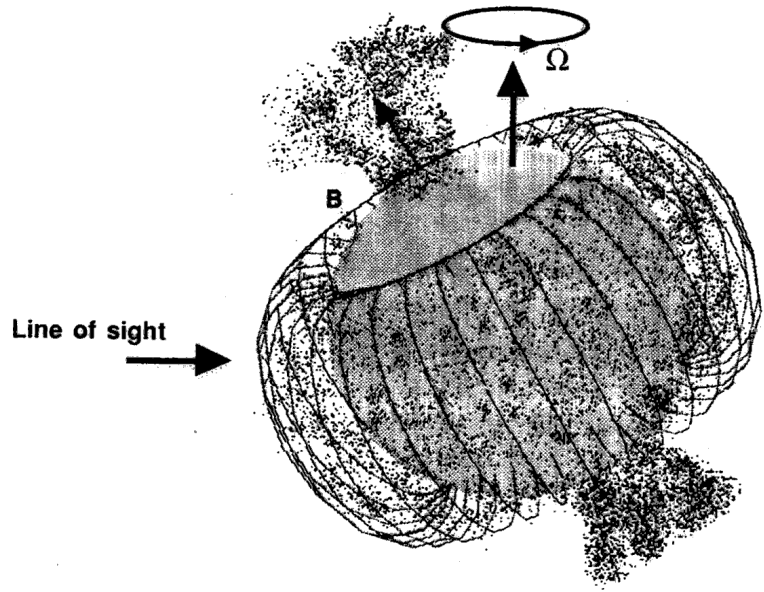}}
	\caption{A cartoon illustrating the geometry of the proposed model for an oblique rotator of intermediate obliquity \citep{Shore1990}}
	\label{fig:8}
\end{figure}

The model of \cite{Shore1990} is based on the observed correlations between the line profiles characteristics of the C\,IV and Si\,IV ultraviolet resonance doublet, the variations in the strength of these lines, and the inferred magnetic field geometry in helium stars (incidentally, the atmosphere of the donor also has excess helium). In Fig.~\ref{fig:8} (representing Figure 11b of Shore and Brown (1990)) depicts the model of a helium star as an oblique magnetic rotator. This model may be to some extent illustrative of the donor as an oblique magnetic rotator since the dipole axis of its magnetic field is deviated by 28$\degr$ relative to the orbital plane of the binary system \citep{Skulskij1985}. Given that the center of the magnetic field dipole of the donor is substantially displaced in the direction of the gainer, additional outflows of matter from the donor surface in the direction of the magnetic axis of the donor may show an unusual, but the predictable picture of the mass transfer. This picture, although complicated by real magnetohydrodynamics, must be reflected in some way in both the accretionary structures between the components (that is investigated in this article) and the circumbinary structures surrounding the Beta Lyrae system (it is a task of the next article). 

\section{Conclusions and discussion}
Throughout the present investigation, it has been confirmed that two types of magnetic field studies by \cite{Skulskij1985, Skulskij1990}, based on observations of the 6-m SAO telescope; and \cite{Skulsky1993} based on CCD observations of the 2.6-m CrAO telescope), show reasonable expectations in the certain phase correlation with each other. Although their averaged absolute values are different, their phase curves of the magnetic field are similar on the whole: they capture the direction of the magnetic field dipole and phases of visibility of the two poles on the donor surface. This made it possible to more reliably study various physical processes reflected in their phase variability parallel with the corresponding variability of the donor magnetic field. In \cite{Skulskyy2020} it is shown that a quasi-sinusoidal photographic curve of the magnetic field, which is similar to such curves of magnetic stars, is more appropriate for this type of research. In the previous sections this approach is worked out on the basis of a number of obvious facts obtained from the analysis of various spectral and spectrophotometric observations mainly in the visible spectrum of this binary system. The focus was on the existence of correlations between the orbital variability of the magnetic field and the analogous variability of certain parameters in different physical processes, primarily due to studying the dynamic and energy characteristics of complex emission-absorption lines. First of all, the variable structure of gas flows in the processes of accretion and mass transfer between components of the binary system and the reflection of such magnetized spatial structures in the orbital variability of the magnetic field was studied. The following are some of the key points of this study as a certain phenomenon.
 
No doubt, the important result was the obtaining of the curve of the phase variability of the absolute radiation flux in the H$_\alpha$ emission line as a certain fact in its evident interconnection with the phase variability of the magnetic field of the donor (see Fig.~\ref{fig:1} in section \ref{sec:2_1}). Firstly, the curve of the variability of the absolute radiation flux in the H$_\alpha$ line clearly displays the direction of (0.355-0.855)\,P of the donor magnetic field maxima and its dipole axis. Secondly, between the extrema of the magnetic field, the radiation flux in the H$_\alpha$ line increases 1.3 times: this flux has a minimum level in the phases near 0.355\,P of the first quadrature, i.e., in phases of the observation of the magnetic field pole on the donor surface, and the maximum level in phases of the second quadrature round the 0.855\,P phase, i.e., in phases of the observation on the donor surface of the magnetic field pole facing the gainer. Thirdly, the apparent additional radiation flux in the H$_\alpha$ line determines the center and the phase limits (0.43 $\pm$ 0.06)\,P of the known hot region of radiating medium projected onto the accretion disk. It is most certain that the effect obtained is the result of a direct collision of gaseous flows with the accretion disk. All in all, one can conclude that the magnetic field is directly reflected in the radiating accretion structures formed during the matter transfer in the Beta Lyrae system.

The above-giving finding is supported by the high-speed variability of the Beta Lyrae spectrum in the region of the H$_\alpha$ line (see section \ref{sec:2_2}). An unusually broad base emission component, with a total width of more than 6000 km/s under a known emission peak with a total width of up to 700 km/s, was detected. This broad emission has the highest continuum in the phases of two quadratures and near phases of the donor magnetic field poles. Such spread wings of the broad emission indicate high speeds of the radiating atoms and reflect directions of the observation of the more high-temperature hydrogen medium. Simultaneously, in these phases, there is the most chaotic variability in the shape and the intensity of the entire emission on 10-second spectra, in both the central peak of the H$_\alpha$ emission line and its broad emission component. In vision phases of the magnetic field pole facing the gainer, concretely at phase 0.81\,P, there was recorded an event of the eruptive nature, which is formed by two components: \textquote{emission flare-up} and \textquote{traveling absorption}. This may reflect a high-velocity matter outflow or even gas jets from a binary system but, that is important, this event took place in the neighborhood of the donor magnetic field pole. Indeed, the characteristic time of the event indicates the local spatial and temporal characteristics of the movement of gaseous structures close to the donor surface. 

This conclusion is supported by other energetics parameters. The phase dependence of the intensity ratio of the violet and red emission peaks of $I_v/I_r=f(P)$ in the He\,I $\lambda$ 7065 emission line reflects two extrema on the curve of the effective magnetic field at the phases 0.355\,P and 0.855\,P, which correspond to the visibility phases of the donor magnetic poles (see Fig.~\ref{fig:6}b). This is the only emission line of the visual spectrum in which its violet peak exceeds the red peak in the 0.355\,P phase, indicating additional radiation directed at the observer. The red peak, as in the H$_\alpha$ emission line, dominates in the 0.855\,P phase, i.e., in the phase of the visibility of the opposite donor magnetic pole facing the gainer. Hence, the dependence of $I_v/I_r=f(P)$,which reflects the radiation gas structures near the donor, is modulated by the spatial configuration of the donor’s magnetic field. The study of the intensity and equivalent width of the various lines formed in both the conditions of the donor peculiar atmosphere reaching its Roche cavity and the immediate vicinity of the donor \ref{sec:2_4}) shows that the donor magnetic field significantly influences their behavior. For example, a clear correlation was found between the phase variations of the equivalent width of Si\,II-emissions and such dependence $W_\lambda = f(P)$ for absorption lines of these complex Si\,II $\lambda \lambda$ 6347, 6371 lines in the donor's atmosphere. There is an apparent modulation of these curves by the donor magnetic field. This also can be seen from gradual changes in the profiles of lines of the Balmer series from their first to the highest members of the series. The most clearly pronounced stratification in physical conditions in the donor atmosphere and above its surface is observed in the phases of (0.85 $\pm$ 0.15)\,P, i.e., in phases of the visibility of the donor magnetic pole facing the gainer. Thus, the aforementioned and other factors indicate this important phase region of the matter loss from the deformed donor surface. However, only the analysis of the light curves at M band (4.6 $\mu$m) and others (see section \ref{sec:2_5}) allowed the direct identification on the donor surface of the magnetic pole facing the gainer. The effective width of its phase visibility is near 0.1\,P and its center matches the 0.855P phase of the magnetic pole center. 

Parallel with the above energetics parameters, the correlation ligaments of the effective magnetic field strength of the donor in its variability over the orbital phases with the dynamics characteristics of the H$_\alpha$ emission line and other known lines of visual spectrum are equally compelling. Three variable structural components are traditionally measured: two emission peaks and the absorption feature between these emission peaks. The measurement of the Doppler shifts of the components of these emission lines showed that the new parameter we introduced \citep{Skulskii1992-Malkov}, namely the Gaussian emission center as a whole, became the most informational. Subsequent observations and reliable measurements of the high-resolution CCD spectra \citep{Skulskij1993b} confirmed that the radial velocity curves of such centers of emission profiles (primarily H$_\alpha$ and He\,I $\lambda$ 7065) fully correspond to the orbital curve of the magnetic donor field variability (see section \ref{sec:2_3}). Indeed, the radial velocity of the center of this total emission (see Fig.~\ref{fig:5} and \ref{fig:6}) shows that both maxima clearly match the two extrema of the sinusoidal curve of the effective magnetic field strength of the donor (see Fig.~\ref{fig:1}). This indicates that these two curves reflect physically related phenomena. It should be supposed that the emission as a whole of complex emission-absorption lines is essentially formed under the influence of the spatial structure of the donor magnetic field, which is reflecting by the (0.355-0.855)\,P phases of its dipole axis. Moreover, the dynamical characteristics of the Gaussian profile of H$_\alpha$ emission as a whole fully match the findings of the orbital variability of the absolute radiation flux in the $H_\alpha$ emission line (see the above and section \ref{sec:2_1}), i.e., the dynamic and energy characteristics of this emission are formed to a large extent in magnetized structures near the donor. This is an important conclusion.

The second important finding follows also from measuring these Doppler shifts: the absorption feature in the limits of the total double-peaked emission is the result of self-absorption in magnetized structures surrounding the donor. This result became better clear after comparing the radial velocity curves of the absorption cores of all the studied complex lines, primarily H$_\alpha$ and He\,I $\lambda$ 7065. This can be seen in Fig.~\ref{fig:6}c, where the absorption component of the He\,I $\lambda$ 7065 line reaches positive velocities of +15 km/s in phases (0.35-0.55)\,P (that is, the movement of absorption material flowing off from the donor surface in the direction from the observer), while for the H$_\alpha$ line and the rest of the lines this component has a known minimum of -15 km/s in the phase of the accretion disk eclipse by the donor (the motion of this matter to the observer). It means that dynamic parameters of the radiating and absorption medium, in which these lines are formed, differ in both density and excitation conditions; i.e., such medium is substantially stratified. 

There are some other points as to dynamics and the spatial formation of the radiating-absorption gaseous structures. Figure~\ref{fig:6}c demonstrates that the absorption components of all the studied lines reach a known maximum negative radial velocity of -115 km/s (close to parabolic) in the direction of the gainer in phases of about 0.05\,P. This can be explained by the action of Coriolis force that form the main gas flow within the Roche cavity of the gainer. Such flow does not leave the limits of the gainer gravity and is important for the formation of an accretion disk around the gainer (see \cite{Skulskyy2020}). It is important to note again the direction of motion of the radiating plasma along the donor axis. Figure~\ref{fig:6}c shows that positive radial velocities of +15 km/s of the absorption component on the He\,I $\lambda$ 7065 emission line, closely 0.355\,P phases of the observation of the donor magnetic pole, reflect, in fact, the direction of the mass loss from the donor surface toward the opposite donor magnetic pole facing the gainer. From Fig.~\ref{fig:5}d it is well discernible that the violet peak of all the studied lines reaches the maximum value of the negative radial velocity of -200 km/s (to the observer) in the phase of 0.855\,P of the observation of the donor magnetic pole facing the gainer. Besides, the radial velocity curve for the more intense red emission peak in Fig.~\ref{fig:5}a has two maxima of the positive radial velocity at phases of the observation of both donor magnetic poles. In general, all this can be interpreted as the matter outflow from the region of the magnetic poles of the donor surface along the direction of (0.355-0.855)\,P of the magnetic field axis. In the process of the formation of moving accretionary structures, the decisive role is played by the donor magnetic pole facing the gainer.

 Thus, the above-giving specification of the main results, presented in the form of some research discussion, suggests the concept of formation of developed spatial accretion structures in the Beta Lyrae system in the presence of the donor magnetic field. It can be summarized as follows.

The analysis of a number of spectral and spectrophotometric observations showed that the structure of the gaseous flows between the donor and the gainer is largely due to the specific dipole configuration of the donor magnetic field. The axis of the donor magnetic field, which reflects the two extrema of the magnetic field strength curve, there is in the direction in the orbital phases of (0.355-0.855)\,P and is deflected from the phases axis of (0.5-1.0)\,P, i.e., from the direction of the gravitational axis of the stellar components. The magnetic field axis is inclined to the orbital plane by an angle of 28$\degr$. The center of the donor magnetic dipole is displaced by 0.08 of the distance between the centers of gravity of both components toward the gainer center. The magnetic pole, which is observed in the phases near 0.855\,P, is located on the donor surface slightly above the orbital plane and closer to the gainer. The localization of this magnetic pole is important in the context of a few next issues. The ionized gas channeled by the donor magnetic field moves in the direction of its dipole axis from the donor surface and deflects along the magnetic field lines toward the accretion disk. In the space between the donor and the gainer during the matter motion, there is formed a system of developed magnetized radiating flows, which is clearly reflected from the study of their physical characteristics, especially in the second quadrature. Effective shock collisions of the magnetized plasma in the phases of this second quadrature of (0.6-0.8)\,P are enhanced by the rapid counter-rotation of the accretion disk with the formation of a high-temperature medium and a hot arc on its outer rim facing the donor. In the plane with the axis of the magnetic field, there is a significant vertical component of the flux of the magnetic induction vector, which can be responsible for the gaseous jets observed as perpendicular to the plane of the orbit.
  
It should be noted that the obtained picture characterizes, first of all, the accretion structures in the space between the two stellar components. A more general picture must be created from the study of expanding structures surrounding this binary system. Some points of such study are published in \cite{Skulsky2015}. More thorough research is planned to be published in the next article.

\acknowledgements
The author is thankful to 
Kudak V.I. for consultations.

\bibliography{pp717-747}

\begin{thebibliography}{34}
\expandafter\ifx\csname natexlab\endcsname\relax\def\natexlab#1{#1}\fi

\bibitem[{{Alexeev} \& {Skulskij}(1989)}]{Alekseev1989}
{Alexeev}, G.~N. \& {Skulskij}, M.~Y., {Rapid variability of the spectrum of
  {\ensuremath{\beta}} Lyrae in the {\ensuremath{H_\alpha}} region}. 1989, {\it
  Bull. Spec. Astroph. Obs.}, {\bf 28}, 21

\bibitem[{{Appenzeller} \& {Hiltner}(1967)}]{Appenzeller1967}
{Appenzeller}, I. \& {Hiltner}, W.~A., {True Polarization Curves for Beta
  Lyrae}. 1967, {\it \apj}, {\bf 149}, 353, DOI: 10.1086/149258

\bibitem[{{Aydin} {et~al.}(1988){Aydin}, {Brandi}, {Engin}, {Ferrer}, {Hack},
  {Sahade}, {Solivella}, \& {Yilmaz}}]{Aydin1988}
{Aydin}, C., {Brandi}, E., {Engin}, S., {et~al.}, {A sudy of the continuum flux
  and the line structure in the IUE spectrum of {\ensuremath{\beta} Lyrae}}.
  1988, {\it \aap}, {\bf 193}, 202

\bibitem[{{Bahyl}(1986)}]{Bahyl1986}
{Bahyl}, V., {Equivalent Widths of the Spectral Lines of the
  {\ensuremath{\beta}} Lyrae System and Their Changes}. 1986, {\it \bac}, {\bf
  37}, 42

\bibitem[{{Batten} \& {Sahade}(1973)}]{Batten1973}
{Batten}, A.~H. \& {Sahade}, J., {The Emission Profile of
  H{\ensuremath{\alpha}} in the Spectrum of {\ensuremath{\beta}} Lyrae}. 1973,
  {\it \pasp}, {\bf 85}, 599, DOI: 10.1086/129511

\bibitem[{{Bisikalo} {et~al.}(2000){Bisikalo}, {Harmanec}, {Boyarchuk},
  {Kuznetsov}, \& {Hadrava}}]{Bisikalo2000}
{Bisikalo}, D.~V., {Harmanec}, P., {Boyarchuk}, A.~A., {Kuznetsov}, O.~A., \&
  {Hadrava}, P., {Circumstellar structures in the eclipsing binary eta Lyr A.
  Gasdynamical modelling confronted with observations}. 2000, {\it \aap}, {\bf
  353}, 1009

\bibitem[{{Bless} {et~al.}(1976){Bless}, {Eaton}, \& {Meade}}]{Bless1976}
{Bless}, R.~C., {Eaton}, J.~A., \& {Meade}, M.~R., {Random variations in the
  ultraviolet spectrum of {\ensuremath{\beta}} Lyrae.} 1976, {\it \pasp}, {\bf
  88}, 899, DOI: 10.1086/130043

\bibitem[{{Burnashev} \& {Skulskij}(1986)}]{Burnashev1986}
{Burnashev}, V.~I. \& {Skulskij}, M.~Y., {Variability of the 5200\,A in
  {\ensuremath{\beta}} Lyrae and {\ensuremath{\nu}} Sagittarii.} 1986, {\it
  Pisma v Astronomicheskii Zhurnal}, {\bf 12}, 535

\bibitem[{{Burnashev} \& {Skulskij}(1991)}]{Burnashev1991}
{Burnashev}, V.~I. \& {Skulskij}, M.~Y., {H$_{{\ensuremath{\alpha}}}$
  photometry and magnetic field of {\ensuremath{\beta}} lyrae}. 1991, {\it
  \krym}, {\bf 83}, 108

\bibitem[{{Harmanec} {et~al.}(1996){Harmanec}, {Morand}, {Bonneau}, {Jiang},
  {Yang}, {Guinan}, {Hall}, {Mourard}, {Hadrava}, {Bozic}, {Sterken},
  {Tallon-Bosc}, {Walker}, {McCook}, {Vakili}, {Stee}, \& {Le
  Contel}}]{Harmanec1996}
{Harmanec}, P., {Morand}, F., {Bonneau}, D., {et~al.}, {Jet-like structures in
  {\ensuremath{\beta}} Lyrae. Results of optical interferometry, spectroscopy
  and photometry.} 1996, {\it \aap}, {\bf 312}, 879

\bibitem[{{Hoffman} {et~al.}(1998){Hoffman}, {Nordsieck}, \&
  {Fox}}]{Hoffman1998}
{Hoffman}, J.~L., {Nordsieck}, K.~H., \& {Fox}, G.~K., {Spectropolarimetric
  Evidence for a Bipolar Flow in beta Lyrae}. 1998, {\it \aj}, {\bf 115}, 1576,
  DOI: 10.1086/300274

\bibitem[{{Ignace} {et~al.}(2018){Ignace}, {Gray}, {Magno}, {Henson}, \&
  {Massa}}]{Ignace2018}
{Ignace}, R., {Gray}, S.~K., {Magno}, M.~A., {Henson}, G.~D., \& {Massa}, D.,
  {A Study of H{\ensuremath{\alpha}} Line Profile Variations in
  {\ensuremath{\beta}} Lyr}. 2018, {\it \aj}, {\bf 156}, 97, DOI:
  10.3847/1538-3881/aad339

\bibitem[{{Jameson} \& {Longmore}(1976)}]{Jameson1976}
{Jameson}, R.~F. \& {Longmore}, A.~J., {Infrared observations and a model for
  {\ensuremath{\beta}} Lyr.} 1976, {\it \mnras}, {\bf 174}, 217, DOI:
  10.1093/mnras/174.1.217

\bibitem[{{Kondo} {et~al.}(1994){Kondo}, {McCluskey}, {Silvis}, {Polidan},
  {McCluskey}, \& {Eaton}}]{Kondo1994}
{Kondo}, Y., {McCluskey}, G.~E., {Silvis}, J. M.~S., {et~al.}, {Ultraviolet
  Light Curves of beta Lyrae: Comparison of OAO A-2, IUE, and Voyager
  Observations}. 1994, {\it \apj}, {\bf 421}, 787, DOI: 10.1086/173691

\bibitem[{{Lomax} {et~al.}(2012){Lomax}, {Hoffman}, {Elias}, {Bastien}, \&
  {Holenstein}}]{Lomax2012}
{Lomax}, J.~R., {Hoffman}, J.~L., {Elias}, Nicholas~M., I., {Bastien}, F.~A.,
  \& {Holenstein}, B.~D., {Geometrical Constraints on the Hot Spot in Beta
  Lyrae}. 2012, {\it \apj}, {\bf 750}, 59, DOI: 10.1088/0004-637X/750/1/59

\bibitem[{{Mennickent} \& {Djura{\v{s}}evi{\'c}}(2013)}]{Mennickent2013}
{Mennickent}, R.~E. \& {Djura{\v{s}}evi{\'c}}, G., {On the accretion disc and
  evolutionary stage of {\ensuremath{\beta}} Lyrae}. 2013, {\it \mnras}, {\bf
  432}, 799, DOI: 10.1093/mnras/stt515

\bibitem[{{Sahade} {et~al.}(1959){Sahade}, {Huang}, {Struve}, \&
  {Zebergs}}]{Sahade1959}
{Sahade}, J., {Huang}, S.~S., {Struve}, O., \& {Zebergs}, V., The Spectrum of
  Beta Lyrae. 1959, {\it Transactions of the American Philosophical Society},
  {\bf 49}, 1

\bibitem[{{Shore} \& {Brown}(1990)}]{Shore1990}
{Shore}, S.~N. \& {Brown}, D.~N., {Magnetically Controlled Circumstellar Matter
  in the Helium-strong Stars}. 1990, {\it \apj}, {\bf 365}, 665, DOI:
  10.1086/169520

\bibitem[{{Skulskij}(1972)}]{Skulskij1972}
{Skulskij}, M.~Y., {Quantitative analysis of {\ensuremath{\beta}} Lyrae
  spectra. I. Variations of some hydrogen and helium lines.} 1972, {\it \krym},
  {\bf 45}, 135

\bibitem[{{Skulskij}(1980)}]{Skulskij1980_PAZ}
{Skulskij}, M.~Y., {Short-term fluctuations of hydrogen and helium emission in
  the spectrum of {\ensuremath{\beta}} Lyrae}. 1980, {\it Pisma v
  Astronomicheskii Zhurnal}, {\bf 6}, 628

\bibitem[{{Skulskij}(1982)}]{Skulskij1982}
{Skulskij}, M.~Y., {Beta Lyrae as a magnetic binary star}. 1982, {\it \sal},
  {\bf 8}, 126

\bibitem[{{Skulskij}(1985)}]{Skulskij1985}
{Skulskij}, M.~Y., {The Magnetic Field of the Beta-Lyrae System}. 1985, {\it
  \sal}, {\bf 11}, 21

\bibitem[{{Skulskij}(1990)}]{Skulskij1990}
{Skulskij}, M.~Y., {The magnetic field and tidal resonant phenomena in
  {\ensuremath{\beta}} Lyrae.} 1990, {\it Mittelungen KSO Tautenberg}, {\bf
  125}, 146

\bibitem[{Skulskij(1992)}]{Skulskij1992}
Skulskij, M.~Y., Study of {\ensuremath{\beta}} Lyrae CCD spectra. Absorbtion
  lines, orbital elements and disk structure of the gainer. 1992, {\it \sal},
  {\bf 18}, 287

\bibitem[{{Skulskij}(1993{\natexlab{a}})}]{Skulskij1993b}
{Skulskij}, M.~Y., {Spectra of {\ensuremath{\beta}} Lyr. Matter transfer and
  circumstellar structures in presence of the donor’s magnetic field}.
  1993{\natexlab{a}}, {\it Astron. Lett.}, {\bf 19}, 45

\bibitem[{{Skulskij}(1993{\natexlab{b}})}]{Skulskij1993a}
{Skulskij}, M.~Y., {Spectra of {\ensuremath{\beta}} Lyr the SiII
  {\ensuremath{\lambda \lambda}}\,6347, 6371 doublet and the cyclic variation
  of the equivalent widths of lines in the "magnitized" atmosphere of the
  loser}. 1993{\natexlab{b}}, {\it Astron. Lett.}, {\bf 19}, 19

\bibitem[{{Skulskij}(1993{\natexlab{c}})}]{Skulskij1993c}
{Skulskij}, M.~Y., {The spectrum of {\ensuremath{\beta}} Lyrae: the SiII
  {\ensuremath{\lambda \lambda}}\,6347, 6371 doublet in 1992 and its variation
  from season to season}. 1993{\natexlab{c}}, {\it Astron. Lett.}, {\bf 19},
  160

\bibitem[{{Skulskij} \& {Malkov}(1992)}]{Skulskii1992-Malkov}
{Skulskij}, M.~Y. \& {Malkov}, Y.~F., {Investigation of {\ensuremath{\beta}}
  Lyrae based on high-dispersion CCD spectrograms near the
  H{\ensuremath{\alpha}} line}. 1992, {\it \sovast}, {\bf 36}, 147

\bibitem[{{Skulskij} \& {Plachinda}(1993)}]{Skulsky1993}
{Skulskij}, M.~Y. \& {Plachinda}, S.~I., {A study of the magnetic field of the
  bright component of {\ensuremath{\beta}} Lyr in the SiII {\ensuremath{\lambda
  \lambda}}\,6347, 6371 lines}. 1993, {\it Pisma Astron. Zh.}, {\bf 19}, 517

\bibitem[{{Skulskij} \& {Vovchik}(1971)}]{Skulskij1971}
{Skulskij}, M.~Y. \& {Vovchik}, E.~B., {The equivalent widths of the absorption
  lines in the spectrum of beta Lyrae.} 1971, {\it Tsirk. Astron. Obs. Lviv},
  {\bf 45}, 25

\bibitem[{{Skulsky}(2015)}]{Skulsky2015}
{Skulsky}, M.~Y., {On the nature of the interacting {\ensuremath{\beta}} Lyrae
  system: location of hot region on accretion disk as representation of the
  magnetized gas structures}. 2015, {\it Science and Education a New Dimension.
  Natural and Technical Sciences}, {\bf III(6)}, 6

\bibitem[{{Skulsky}(2018)}]{Skulsky2018}
{Skulsky}, M.~Y., {On the structure of magnetized accretion flows in the system
  of Beta Lyrae}. 2018, {\it \caos}, {\bf 48}, 300

\bibitem[{{Skulskyy}(2020)}]{Skulskyy2020}
{Skulskyy}, M.~Y., {Formation of magnetized spatial structures in the Beta
  Lyrae system. I. Observation as a research background of this phenomenon.}
  2020, {\it \caos}, {\bf 50}, 681

\bibitem[{{Zeilik} {et~al.}(1982){Zeilik}, {Heckert}, {Henson}, \&
  {Smith}}]{Zeilik1982}
{Zeilik}, M., {Heckert}, P., {Henson}, G., \& {Smith}, P., {Infrared photometry
  of beta Lyrae: 1977-1982.} 1982, {\it \aj}, {\bf 87}, 1304, DOI:
  10.1086/113217

\end{thebibliography}

\end{document}